\documentclass[prc,twocolumn,showpacs,preprintnumbers,amsmath,amssymb]{revtex4-1}
\usepackage[dvips]{graphicx}
\usepackage{dcolumn}
\usepackage{bm}

\begin{document}
\title{Effects of ground-state correlations on collective excitations of $^{16}$O}

\author{Mitsuru Tohyama}
\affiliation{Kyorin University School of Medicine, Mitaka, Tokyo
  181-8611, Japan} 
\begin{abstract}
The effects of the correlations in the ground state of $^{16}$O on the octupole and dipole excitations are studied using the extended
random phase approximation (ERPA) derived from the time-dependent density-matrix theory. 
It is found that the ground-state correlation effects are significant especially in the octupole excitation. It is shown that
the first $3^-$ state calculated in the random phase approximation (RPA) is shifted upward when the self-energy contributions
are included in particle - hole pairs. 
The coupling to the two particle - two hole states plays a role in shifting the first $3^-$ state down to the right position.
It is also found that the dipole strength is fragmented due to the partial occupation of the single-particle states and 
that the peak position of the giant dipole resonance calculated in ERPA
is little changed from that in RPA due to the above-mentioned competing effects: the increase in particle - hole energy 
and the coupling to two particle - two hole configurations.
\end{abstract}
\maketitle
The random phase approximation (RPA) based on the Hartree-Fock (HF) ground state has extensively been used to study nuclear collective
excitations. It is generally considered that the HF + RPA approach is the most appropriate for double-closed shell nuclei 
for which the HF theory would give a good description of the ground states. However,
recent theoretical studies for $^{16}$O indicate that the ground state of $^{16}$O is a highly correlated state \cite{Toh,utsuno},
which indicates the necessity of using beyond RPA theories to study collective excitations of $^{16}$O. 
In this paper we investigate how octupole and dipole excitations of $^{16}$O are affected by ground-state correlations, using
an extended RPA (ERPA) that has been derived from the time-dependent density-matrix theory (TDDM) \cite{WC,GT}.
We show that the octupole excitation of $^{16}$O is quite sensitive to the ground-state correlation effects.

The TDDM consists of the coupled equations of motion for the one-body density matrix $n_{\alpha\alpha'}$ 
(the occupation matrix) and the correlated part of the two-body density matrix $C_{\alpha\beta\alpha'\beta'}$
(the correlation matrix).
These matrices are defined as
\begin{eqnarray}
n_{\alpha\alpha'}(t)&=&\langle\Phi(t)|a^\dag_{\alpha'} a_\alpha|\Phi(t)\rangle,
\\
C_{\alpha\beta\alpha'\beta'}(t)&=&\langle\Phi(t)|a^\dag_{\alpha'}a^\dag_{\beta'}
 a_{\beta}a_{\alpha}|\Phi(t)\rangle
\nonumber \\
 &-&(n_{\alpha\alpha'}(t)n_{\beta\beta'}(t)
 -n_{\alpha\beta'}(t)n_{\alpha\beta'}(t)),
\end{eqnarray}
where $|\Phi(t)\rangle$ is the time-dependent total wavefunction
$|\Phi(t)\rangle=\exp[-iHt] |\Phi(t=0)\rangle$. The equations of motion for reduced density matrices form
a chain of coupled equations known as the Bogoliubov-Born-Green-Kirkwood-Yvon (BBGKY) hierarchy. In TDDM the BBGKY
hierarchy is truncated by replacing a three-body density matrix with anti-symmetrized products of the one-body and
two-body density matrices. The TDDM equation for $C_{\alpha\beta\alpha'\beta'}$ contains all effects of two-body correlations;
particle - particle, hole - hole and particle - hole correlations.
The ground state in TDDM is given as a stationary solution of the TDDM equations. 
The stationary solution can be obtained using the gradient method \cite{Toh}.
This method is also used in the present work.
The ERPA equations used here are derived as the small amplitude limit of TDDM and are written in matrix form
for the one-body and two-body amplitudes $x^\mu_{\alpha\alpha'}$ and $X^\mu_{\alpha\beta\alpha'\beta'}$ \cite{Toh}
\begin{eqnarray}
\left(
\begin{array}{cc}
A&C\\
B&D
\end{array}
\right)\left(
\begin{array}{c}
x^\mu\\
X^\mu
\end{array}
\right)
=\omega_\mu
\left(
\begin{array}{cc}
S_{1}&T_{1}\\
T_{2}&S_{2}
\end{array}
\right)
\left(
\begin{array}{c}
x^\mu\\
X^\mu
\end{array}
\right).
\label{ERPA}
\end{eqnarray}
The one-body sector of Eq. (\ref{ERPA}) $Ax^\mu=\omega_\mu S_{1}x^\mu$ is formally the same as the equation
in the self-consistent RPA (SCRPA) \cite{Janssen}, which includes the effects of ground-state correlations through
$n_{\alpha\alpha'}$ and $C_{\alpha\beta\alpha'\beta'}$. 
We refer to the approximation $Ax^\mu=\omega_\mu S_{1}x^\mu$ as the modified RPA (mRPA).
To explain the role of the correlation matrix in the mRPA equation, we explicitly show
the matrices $A$ and $S_{1}$:
\begin{eqnarray}
A&(&\alpha\alpha':\lambda\lambda')=
(\epsilon_\alpha-\epsilon_{\alpha'})
(n_{\alpha'\alpha'}
-n_{\alpha\alpha})\delta_{\alpha\lambda}\delta_{\alpha'\lambda'}
\nonumber \\
&+&(n_{\alpha'\alpha'}-n_{\alpha\alpha})(n_{\lambda'\lambda'}-n_{\lambda\lambda})\langle\alpha\lambda'|v|\alpha'\lambda\rangle_A
\nonumber \\
&-&\delta_{\alpha'\lambda'}\sum_{\gamma\gamma'\gamma''}\langle\alpha\gamma|v|\gamma'\gamma''\rangle
C_{\gamma'\gamma''\lambda\gamma}
\nonumber \\
&-&\delta_{\alpha\lambda}\sum_{\gamma\gamma'\gamma''}\langle\gamma\gamma'|v|\alpha'\gamma''\rangle
C_{\lambda'\gamma''\gamma\gamma'}
\nonumber \\
&+&\sum_{\gamma\gamma'}(\langle\alpha\gamma|v|\lambda\gamma'\rangle_A
C_{\lambda'\gamma'\alpha'\gamma}
+\langle\lambda'\gamma|v|\alpha'\gamma'\rangle_A
C_{\alpha\gamma'\lambda\gamma})
\nonumber \\
&-&\sum_{\gamma\gamma'}(\langle\alpha\lambda'|v|\gamma\gamma'\rangle
C_{\gamma\gamma'\alpha'\lambda}
+\langle\gamma\gamma'|v|\alpha'\lambda\rangle
C_{\alpha\lambda'\gamma\gamma'}),
\label{A-term}
\end{eqnarray}
\begin{eqnarray}
S_{1}(\alpha\alpha':\lambda\lambda')=(n_{\alpha'\alpha'}
-n_{\alpha\alpha})\delta_{\alpha\lambda}\delta_{\alpha'\lambda'},
\label{snorm}
\end{eqnarray}
where the subscript $A$ means that the corresponding matrix is antisymmetrized and $n_{\alpha\alpha'}$ is assumed to be diagonal.
The first two terms on the right-hand side of Eq. (\ref{A-term}) are the same as those in the RPA equation, 
the next two terms with $C_{\alpha\beta\alpha'\beta'}$ describe the 
self-energy of the particle - hole (p-h) state due to ground-state correlations \cite{Janssen}, and
the last four terms with $C_{\alpha\beta\alpha'\beta'}$ may be
interpreted as the modification of the p-h interaction
caused by ground-state correlations \cite{Janssen}.
The difference between mRPA and SCRPA is in the method to determine $n_{\alpha\alpha'}$ and 
$C_{\alpha\beta\alpha'\beta'}$: in SCRPA these matrices are self-consistently determined using the
amplitudes $x^\mu_{\alpha\alpha'}$ whereas here we calculate them in TDDM.
If the correlated ground state is replaced by the HF ground state, mRPA and ERPA
become the same as RPA and the second RPA (SRPA) \cite{srpa}, respectively.

The occupation probability $n_{\alpha\alpha}$ and the correlation matrix $C_{\alpha\beta\alpha'\beta'}$ are
calculated within TDDM using the $1p_{3/2}$, $1p_{1/2}$, $1d_{5/2}$ and $2s_{1/2}$ states for both protons and neutrons.
For the calculations of the single-particle states we use the Skyrme III force.
To reduce the dimension size, we only consider the two particle - two hole (2p-2h) and 2h-2p elements of $C_{\alpha\beta\alpha'\beta'}$.
A simplified interaction which contains only the $t_0$ and $t_3$ terms of the Skyrme III force is used as the residual interaction.
The spin-orbit force and Coulomb interaction are also omitted from the residual interaction.
To avoid a cumbersome treatment of the rearrangement effects of a density-dependent force in extended RPA theories \cite{Grasso},
we use the three-body version of the Skyrme interaction,
$v_3=t_3\delta^3({\bm r_1}-{\bm r_2})\delta^3({\bm r_1}-{\bm r_3})$,
which gives the following density-dependent two-body residual interaction:
$t_3\rho_n\delta^3({\bm r}-{\bm r'})$, $t_3\rho\delta^3({\bm r}-{\bm r'})/2$ and 
$t_3\rho_p\delta^3({\bm r}-{\bm r'})$ for the proton-proton, proton-neutron and neutron-neutron interactions, respectively,
where $\rho_p$, $\rho_n$ and $\rho$ are the proton, neutron and total densities, respectively.
In the RPA, mRPA and ERPA calculations
the one-body amplitudes $x^\mu_{\alpha\alpha'}$ are defined using a large number of single-particle states including those in the 
continuum. We discretize the continuum states by confining the wavefunctions in a sphere with radius 15 fm and take all 
the single-particle states with $\epsilon_\alpha\le 50$ MeV and 
$j_\alpha\le 11/2 \hbar$. As the residual interaction, we use the same simple force as that used in the ground-state
calculation.
Since the residual interaction is not consistent with the effective interaction used in the calculation of the single-particle states,
it is necessary to reduce the strength of the residual interaction so that the spurious mode corresponding
to the center-of-mass motion comes at zero excitation energy in RPA. We found 
the reduction factor $f$ is 0.62. The p-h interaction in the second term on the
right-hand side of Eq. (\ref{A-term}) is multiplied with this $f$.
To define the two-body amplitudes 
$X^\mu_{\alpha\beta\alpha'\beta'}$, we use the small single-particle space consisting of the 
$1s_{1/2}$, $1p_{3/2}$, $1p_{1/2}$, $1d_{5/2}$, $2s_{1/2}$, 
$1d_{3/2}$, $2p_{3/2}$, $2p_{1/2}$ and $1f_{7/2}$ for both protons and neutrons.
To reduce the number of the two-body amplitudes, we consider only  
the 2p-2h and 2h-2p components of $X^\mu_{\alpha\beta\alpha'\beta'}$
with $|\epsilon_\alpha+\epsilon_\beta-\epsilon_{\alpha'}-\epsilon_{\beta'}|\le 60$ MeV. Since the single-particle space used for
$X^\mu_{\alpha\beta\alpha'\beta'}$ is small, we use $f=1$ for the matrix elements of the residual interaction which couple to $X^\mu_{\alpha\beta\alpha'\beta'}$.

\begin{table}
\caption{Single-particle energies $\epsilon_\alpha$ and occupation probabilities 
$n_{\alpha\alpha}$ calculated in TDDM.}
\begin{center}
\begin{tabular}{c|rr|rr} \hline
 &\multicolumn{2}{c|}{$\epsilon_\alpha$ [MeV]}&\multicolumn{2}{c}{$n_{\alpha\alpha}$}\\ \hline 
orbit & proton & neutron  & proton & neutron  \\ \hline
$1p_{3/2}$ & -18.3 & -21.9 & 0.894 & 0.893  \\
$1p_{1/2}$ & -12.3 & -15.7 & 0.868 & 0.865  \\
$1d_{5/2}$ & -3.8 & -7.1 & 0.108 & 0.109   \\ 
$2s_{1/2}$ & 1.1 & -1.6 & 0.019 & 0.021  \\ \hline
\end{tabular}
\label{tab1}
\end{center}
\end{table}

The occupation probabilities calculated in TDDM are shown in Table \ref{tab1}.
The deviation from the HF values ($n_{\alpha\alpha}$=1 or 0) is more than 10\%,
which means that the ground state of $^{16}$O is a strongly correlated state.
A recent shell-model calculation by Utsuno and Chiba \cite{utsuno} also gives a similar 
result for the ground state of $^{16}$O.
The correlation energy $E_c$ in the ground state, which is
defined by $E_c=\sum_{\alpha\beta\alpha'\beta'}\langle\alpha\beta|v|\alpha'\beta'\rangle C_{\alpha'\beta'\alpha\beta}/2$, is $-23.7$ MeV.
A large portion of the correlation energy is compensated by the increase in the mean-field energy due to the fractional occupation
of the single-particle states. The resulting energy gain due to the ground-state correlations, which is given
by the total energy difference between HF and TDDM, is 6.4 MeV, which is much smaller than $|E_c|=23.7$ MeV. 
\begin{figure} 
\begin{center} 
\includegraphics[height=5cm]{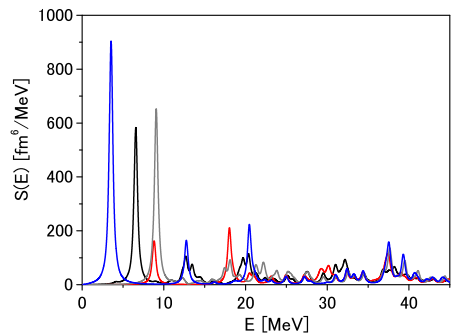}
\end{center}
\caption{(Color online) Strength functions calculated in RPA (blue), SRPA (red), mRPA (grey) and ERPA (black) for the octupole excitation in $^{16}$O. 
The distributions are smoothed with an artificial width $\Gamma=0.5$ MeV.} 
\label{e3} 
\end{figure} 
\begin{figure} 
\begin{center} 
\includegraphics[height=5cm]{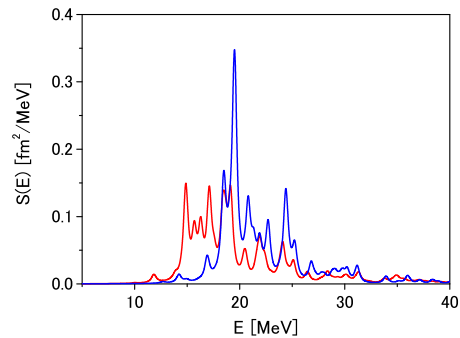}
\end{center}
\caption{(Color online) Strength functions calculated in RPA (blue) and SRPA (red) for the isovector dipole excitation in $^{16}$O. 
The distributions are smoothed with an artificial width $\Gamma=0.5$ MeV.} 
\label{e11} 
\end{figure}
\begin{figure}  
\begin{center} 
\includegraphics[height=5cm]{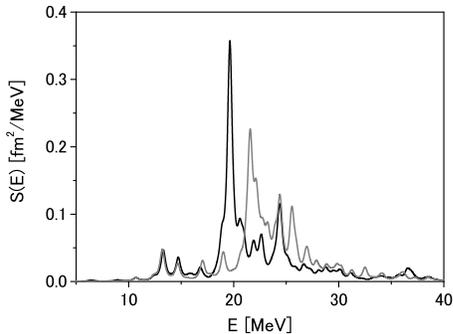}
\end{center}
\caption{Strength functions calculated in mRPA (grey) and ERPA (black) for the isovector dipole excitation in $^{16}$O. 
The distributions are smoothed with an artificial width $\Gamma=0.5$ MeV.} 
\label{e12} 
\end{figure} 

The strength functions for the isoscalar octupole excitation calculated in RPA (blue), SRPA (red), mRPA (grey) and ERPA (black) 
are shown in Fig. \ref{e3}. The excitation operator used is $r^3Y_{30}(\theta)$.
In mRPA and ERPA the fractional occupation of the $2s_{1/2}$ states makes the transitions from these states to the
continuum states possible, which gives some unphysical strength distributions in very low-energy region. As will be shown below,
the unphysical contributions could in principle be eliminated by using the correlation matrix $C_{\alpha'\beta'\alpha\beta}$.
However, it is difficult to completely elimination the unphysical states. Therefore, we neglected the single-particle transitions from
the $2s_{1/2}$ states.
In RPA the summed energy-weighted strength exhausts 97\% of the sum rule.
Other approximations give a similar value of the energy-weighted sum of the octupole strength.
The excitation energy of the first $3^-$ ($3^-_1$) state in RPA is $3.54$ MeV, 
which is much smaller than the experimental data $6.13$ MeV \cite{table}.
In SRPA the state corresponding to the $3^-_1$ state becomes an imaginary solution due to the coupling to the 2p-2h configurations.
The results in RPA and SRPA suggest that the aproaches based on the HF ground state cannot give a correct excitation energy
of the $3^-_1$ state. 
The excitation energy of the $3^-_1$ state in mRPA is $9.1$ MeV, which is much larger than that in RPA.
This is explained by the self-energy contributions (see Eq. (\ref{A-term})) in the p-h pairs such as
$(1p_{1/2})^{-1}\times 2d_{5/2}$ and $(1p_{3/2})^{-1}\times 2d_{5/2}$. The self-energy contributions increase the energy of the 
p-h pairs, reflecting the fact that the ground-state energy is lowered by the ground-state correlations.
We found that the last four terms in Eq. (\ref{A-term}), which describe the modifications of the p-h interaction, play a role
in slightly increasing the attractive p-h correlations. 
In ERPA the $3^-_1$ state is shifted downward
to $6.56$ MeV due to the coupling to the 2p-2h configurations.
The $B(E3)$ value of the $3^-$ state at 6.56 MeV calculated in ERPA 
is $130~e^2$fm$^6$, while the experimental value is $204\pm6~e^2$fm$^6$ \cite{table}.
The results in mRPA and ERPA demonstrate that both the ground-state correlations and the coupling to 2p-2h configurations play an important role
in describing the properties of the $3^-_1$ state.

We point out that the $3^-_1$ state in RPA somewhat depends
on the parameters of the Skyrme interactions.
For example, it has been reported that the SkM$^*$ parameter set \cite{skm} 
gives the $3^-_1$ state at $6.06$ MeV with $B(E3)=91.1~e^2$m$^6$ \cite{hagino}
and that a simple Skyrme type interaction with 
$t_0=-1048$ MeVfm$^3$, $t_3=19150$ MeVfm$^6$ and $w_0=95$ MeVfm$^5$ gives the $3^-_1$ state at $6.05$ MeV which exhausts 10.5\%
of EWSR \cite{abada}. We also found that 
the parameter set of the Skyrme III force with $f=0.62$, $x_0=0$ and $w_0=90$ MeVfm$^5$ gives the $3^-_1$ state at $6.12$ MeV with
$B(E3)=211~e^2$fm$^6$ in RPA, which are close to the experimental data $6.13$ MeV and $204\pm6~e^2$fm$^6$ \cite{table}. 
However, this force also induces strong ground state correlations  
and the $3^-_1$ state calculated in ERPA comes at $8.26$ MeV, which is much higher than the data.

The strength functions for the isovector dipole excitation calculated in RPA (blue) and SRPA (red)
are shown in Fig. \ref{e11} and those in mRPA (grey) and ERPA (black) in Fig. \ref{e12}.  
The single-particle transitions from
the partially occupied $2s_{1/2}$ states are neglected in mRPA and ERPA. 
In RPA the summed energy-weighted strength exhausts 84\% of the dipole sum rule including the enhancement term,  
which is given by the $t_1$ and $t_2$ parameters of the Skyrme III force and accounts 26\% of the sum rule value.
To increase the sum of the energy-weighted strength, we need to include the momentum dependent terms of the Skyme force in the
residual interaction and also expand the single-particle space. 
The energy-weighted sums of the dipole strength in the other approximations give a value similar to that in RPA.
Comparison of the result in SRPA with that in RPA shows that the main effect of the coupling to the 2p-2h configurations
is to shift downward the RPA strength distribution. A similar downward shift of the dipole strength has been reported 
in large scale SRPA calculations \cite{papa,gambacurta10}.
In mRPA the largest peak is upwardly shifted to $21.5$ MeV from the position at $19.5$ MeV in RPA.
This upward shift is due to the self-energy contributions mainly
in the $(1p_{3/2})^{-1} - 1d_{5/2}$ pairs. 
In the case of the isovector dipole excitation, the last four terms in Eq. (\ref{A-term}) play a role in slightly reducing the 
repulsive p-h correlations.
The increase in the dipole strength below 15 MeV in mRPA and ERPA, which is consistent with the experiment \cite{ahrens},
is due to the partial occupation of the $1d_{5/2}$ states allowing such low-energy p-h transitions as $1d_{5/2}\rightarrow 2p_{3/2}$.
The strength distribution above 20 MeV in mRPA is about 2 MeV downwardly shifted in ERPA due to the coupling to the
2p-2h configurations. However, there is little difference between the mRPA and ERPA distributions in 
the energy region below 17 MeV. This indicates that the dipole states in the low-energy region weakly couple to the 2p-2h
configurations.
In the case of the isovector dipole excitation the effects of the ground-state correlations play a role in increasing the fragmentation
of the dipole strength in low-energy region. However, in the giant dipole resonance (GDR) region it seems that 
the increase in the energy of the p-h pairs is compensated by a downward shift of the dipole strength due to the coupling
to the 2p-2h configurations. 
The experimental photoabsorption cross section \cite{ahrens} shows a broader GDR distribution than the result in ERPA.
More two-body configurations should be included to improve the ERPA result.

\begin{figure} 
\begin{center} 
\includegraphics[height=4.5cm]{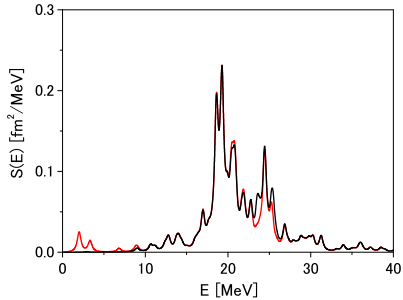}
\end{center}
\caption{(Color online) Strength functions calculated in mRPA (black) and RPA (red) for the isovector dipole excitation in $^{16}$O. 
The partially occupied single-particle states obtained in TDDM are used in the RPA calculation. The correlation matrix
$C_{\alpha'\beta'\alpha\beta}$ is inserted only to the $2s_{1/2}$ - $2p_{3/2}$ and $2s_{1/2}$ - $2p_{1/2}$ pairs
in the mRPA calculation.
The distributions are smoothed with an artificial width $\Gamma=0.5$ MeV.} 
\label{frpae1} 
\end{figure} 

As mentioned above, partial occupation of the particle states could cause some unphysical low-lying transitions possible.
To show this, we performed an RPA calculation for the isovector dipole excitation using
the partially occupied single-particle states
including the $2s_{1/2}$ states. The result is
shown in Fig. \ref{frpae1} with the red line. 
The strength seen below 5 MeV comes from the transitions from the $2s_{1/2}$ states 
to the $2p_{3/2}$ and $2p_{1/2}$ states. The black line in Fig. \ref{frpae1} shows the result in mRPA
where the correlation matrix $C_{\alpha'\beta'\alpha\beta}$ 
is included only in the $2s_{1/2}-2p_{3/2}$ and $2s_{1/2}-2p_{1/2}$ p-h pairs.
Since the norm matrix $S_{1}$ given by Eq. (\ref{snorm}) for the $2s_{1/2}$ - $2p_{3/2}$ and $2s_{1/2}$ - $2p_{1/2}$ pairs is quite small ($\approx0.02$, see Table \ref{tab1}),
the terms in Eq. (\ref{A-term}) which contain $C_{\alpha\beta\alpha'\beta'}$ can drastically shift the energies of these pairs. In fact
the inclusion of the correlation matrix eliminates the strength in the low-energy region and slightly increases
the strength distribution around 25 MeV, as seen in Fig. \ref{frpae1}.
When the coupling of the p-h pairs to 2p-2h configurations is included in ERPA, it becomes difficult to completely eliminate 
the unphysical low-energy transitions from the partial occupied $2s_{1/2}$ states. This is the reason why
we neglected such transitions in the mRPA and ERPA calculations shown above.

In summary, the effects of the correlations in the ground state of $^{16}$O on the octupole and 
dipole excitations were studied using the extended
RPA. It was found that 
the ground-state correlation effects on the first $3^-$ state are significant: the first $3^-$ state is shifted upward
due to the self-energy contributions. It was also found that 
the coupling to two particle - two hole states plays a role in producing the first $3^-$ at right excitation energy.
In the case of the isovector dipole excitation, the effects of the ground-state correlations 
were found to increase the fragmentation
of the dipole strength in low-energy region. However, the giant dipole resonance calculated in ERPA
is little changed from that in RPA due to the competing effects: the increase in particle - hole energy 
and the coupling to two particle - two hole configurations.
The mechanism to eliminate unphysical low-energy transitions originating from partial occupation of the particle states was also discussed.
Our results demonstrate that the ground-state correlation effects in $^{16}$O 
should be properly taken into account in the study of collective excitations.
It is interesting to study these effects in other nuclei.

\end{document}